\documentclass[prl,onecolumn,showpacs,preprintnumbers,amsmath,amssymb,11pt]{revtex4}
\usepackage[utf8]{inputenc}
\usepackage{epsf}
\usepackage{graphicx}
\usepackage{latexsym}
\usepackage{amsmath}
\usepackage{amssymb}
\usepackage{amsfonts}
\usepackage{soul}
\usepackage{color}

\begin{document}


\title{Observation of hidden parts of dislocation loops in thin Pb films \\ by means of scanning tunneling spectroscopy}
\author{A. Yu. Aladyshkin$^{a,b,c}$, A. S. Aladyshkina$^d$ and S. I. Bozhko$^{e}$}
\affiliation{$^{a}$ Institute for Physics of Microstructures RAS, 603950, Nizhny Novgorod, GSP-105, Russia \\
$^{b}$ N.\,I. Lobachevsky State University of Nizhny Novgorod, Nizhny Novgorod, 603022 Russia\\
$^{c}$ HSE University,  Myasnitskaya str. 20, 101000 Moscow, Russia \\
$^{d}$ HSE University, 25/12 Bolshaya Pecherskaya str., Nizhny Novgorod, 603155, Russia\\
$^{e}$ Osipyan Institute of Solid State Physics RAS, Acad. Ossypian str. 2, Chernogolovka, Moscow district, 142432 Russia}


\begin{abstract}
Local electronic properties of quasi-two-dimensional Pb(111) islands with screw dislocations of different types on their surfaces were experimentally studied by means of low-temperature scanning tunneling microscopy and spectroscopy in the regime of constant current. A comparison of the topography map, the maps of tunneling current variation and the differential tunneling conductance acquired simultaneously allows one to visualize the hidden parts of the dislocation loops under the sample surface. We demonstrate that two closely-positioned screw dislocations with the opposite Burgers vectors can either (i) connect to each other by the sub-surface dislocation loop or (ii) generate independent hidden edge dislocation lines which run towards the perimeter of the Pb island. In addition, we found a screw dislocation, which does not produce outcoming sub-surface dislocation loops. Screw dislocations and the hidden dislocations lines are the source of the non-quantized variation of the local thickness of the Pb terraces, affecting the local electronic properties.
\end{abstract}



\maketitle

\section{Introduction}

\begin{figure*}[th!]
\centering{\includegraphics[width=13cm]{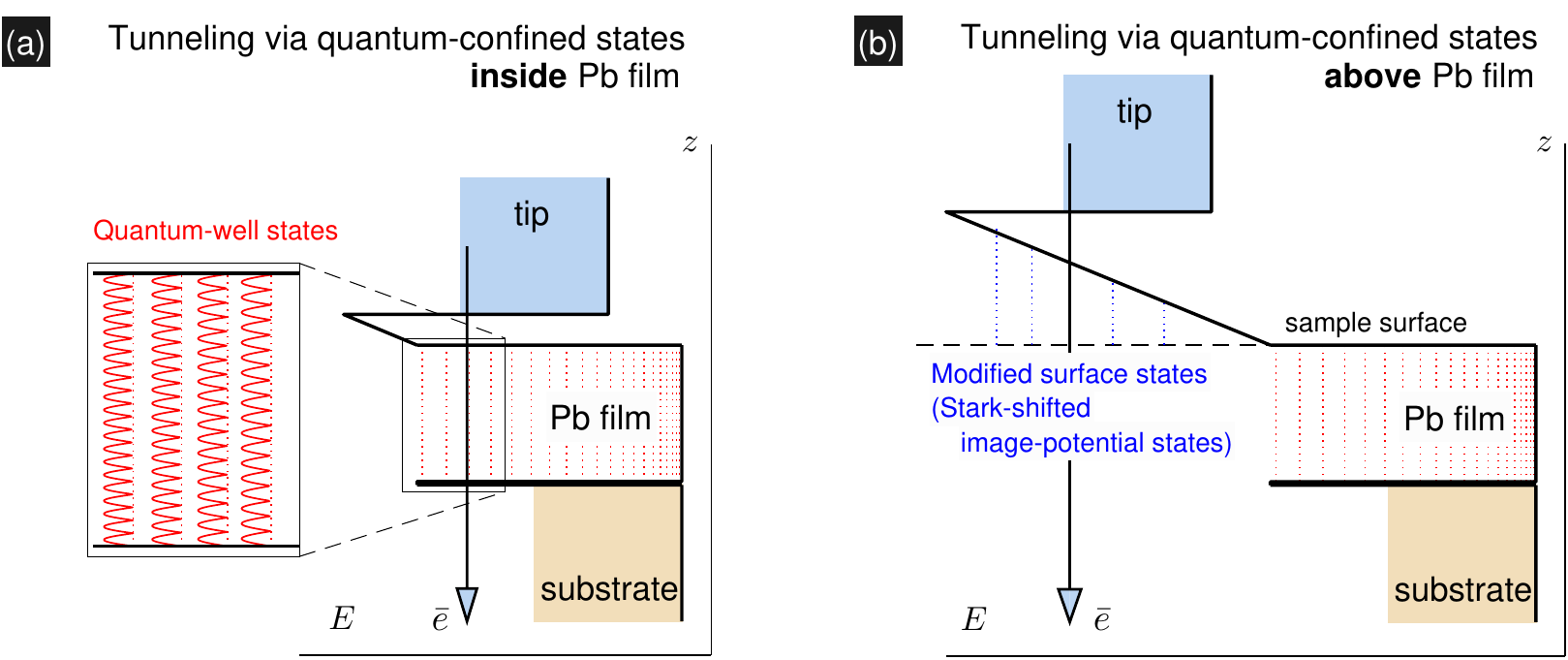}}
\caption{({\bf a, b}) Schematic presentation of the resonant processes of electronic coherent transport from the tip to the sample in a double-barrier quantum system through quasi-stationary electronic states, localized inside the two-dimensional metallic film (a) and above the metallic film (b). Yellow and blue rectangles show the filled electronic states with $E\le E^{\,}_F$ in the sample and the tip, correspondingly; $U=\varphi^{\,}_s-\varphi^{\,}_t>0$. }
\label{Fig01}
\end{figure*}

The formation of discrete quantum-well states (QWSs) caused by a confinement of electrons in micro- and nanostructures is one of the fundamental properties of quantum mechanics (see, \emph{e.g.}, textbooks \cite{Landau-III,Ferry-book-09}). For thin-film samples electrons in the conduction band can be considered as nearly-free particles for motion in the lateral (in-plane) direction and confined for motion in the transverse (out-of-plane) direction resembling a 'particle-in-the-box' problem. For quasi-two-dimensional electronic gas in hybrid systems limited by potential barriers of low transmission the localized solutions of the Schr\"odinger equation are close to standing electron waves with an integer number $n$ of half-waves (see the inset in Fig.~\ref{Fig01}a) and exponentially decaying tails in the barrier area. This apparently leads to a quantization of the wave vector $k^{\,}_{\perp,n}$ and the energy $E^{\,}_{n}$ of the electronic states, confined in the transverse direction. The modification of the energy spectrum is responsible for the formation of series of local maxima in the dependence of a photoemission response on the binding energy for the filled electronic states (see, \emph{e.g.}, reviews \cite{Chiang-SSR-00,Milun-RPP-02} and references therein). Quantum-size effects also modify the dependence of electric conductivity of thin metallic films as well as the Hall coefficient on the thickness \cite{Miyata-PRB-08,Jalochowski-PRB-88}. The effective resonant tunneling \cite{Ferry-book-09} through quasi-stationary QWS (Fig.~\ref{Fig01}a) results in the appearance of almost equidistantly positioned peaks \cite{Comment} on the dependence of differential tunneling conductance on the bias voltage both for the filled and the empty electronic states in experiments involving scanning tunneling microscopy and spectroscopy (STM/STS) \cite{Altfeder-PRL-97,Altfeder-PRL-98,Su-PRL-01,Eom-PRL-06,Hong-PRB-09,Kim-SurfSci-15,Ustavshchikov-JETPLett-17,Putilov-JETPLett-19}. Remarkably, the period of the quantum-size oscillations $\Delta E$ (\emph{i.e.} the energy interval between two neighbor maxima in the photoemission and tunneling experiments) near the Fermi energy depends inversely proportional on the local thickness of the film $D$.

Ultrathin lead (Pb) films and islands appear to be convenient objects for studying quantum-size effects in metallic films in normal and superconducting states by means of transport and Hall measurements \cite{Miyata-PRB-08,Jalochowski-PRB-88} and by STM/STS \cite{Altfeder-PRL-97,Altfeder-PRL-98,Su-PRL-01,Eom-PRL-06,Hong-PRB-09,Kim-SurfSci-15,Ustavshchikov-JETPLett-17,Putilov-JETPLett-19}; for the investigation of electronic properties in superconducting nanostructures \cite{Cren-PRL-09,Ning-ERL-09} and superconducting two-dimensional materials \cite{Brun-SuST-17}, hybrid structures superconductor--ferromagnet \cite{Aladyshkin-PRB-11,Iavarone-NatCom-14}, superconductor--normal metal \cite{Cherkez-PRX-14,Roditchev-NatPhys-15}, and superconductor--topological insulator \cite{Stolyarov-JPCLett-21}; for the investigation of electronic states by means of photoemission electron spectroscopy \cite{Mans-PRB-02}. We would like to emphasize that the STM/STS technique is usually considered as a method for the testing of electronic properties of the surface due to the exponential decay of both STM tip and surface wave functions in a tunnel barrier area. Nevertheless the interference of the electronic wave emanating from the STM tip and the secondary waves scattered by point-like or extended defects under a metal surface give rise to a standing-wave pattern in the lateral plane \cite{Avotina-PRB-05,Weismann-Science-09}. The particular sensitivity of the interference patterns to the variations of the film thickness and the crystalline structure of the interfaces explains why the methods of the tunneling interferometry can be used for the visualization of monatomic steps at the upper/lower interfaces \cite{Altfeder-PRL-97,Altfeder-PRL-98,Kim-SurfSci-15,Ustavshchikov-JETPLett-17}, the atomic lattice of the substrate covered by metal \cite{Altfeder-PRL-98},  inclusions \cite{Ustavshchikov-JETPLett-17}, and terraces with non-quantized height variations \cite{Putilov-JETPLett-19}.

This paper is devoted to the experimental investigations of the spatial inhomogeneity of the differential tunneling conductance near screw and edge dislocations in thin Pb films by means of low--temperature STM/STS. Numerical modeling of thin-film structures and nanostructures with screw dislocations and periodic network of dislocations were presented in Refs. \cite{Nikiforov-JPCLett-11,Zhou-JPCC-12,Olson-JPCLett-16}. Surely, linear defects like dislocations and disclinations at the surface can be easily visualized by scanning tunneling microscopy \cite{Figuera-DislSolids-98,Christiansen-PRL-02,Coupeau-DislSolids-04,Morgenstern-PRB-05,Engbaek-PRB-06,Mansour-NanoLett-08,Dominguez-ChemPhys-10,Bozhko-PRB-14,Weidlich-JAP-15,Zeuthen-JPCC-13} or by atomic-force microscopy \cite{Watkins-CrystGrowth-97,Simpkins-JAP-03,Bennett-RSI-10,Markov-CrystRep-20}. The hidden dislocation lines inside metallic films can be imaged by means of scanning and transmission electron microscopy \cite{Hirsch-65,Crimp-MRT-06,Jin-JPCLett-10,Liu-JPCC-19}. However both atomic-force microscopy and electron microscopy do not allow the testing of local electronic properties (such as local current--voltage dependence, value of superconducting gap etc). It seems interesting to develop experimental methods based on the tunneling interferometry for deeper analysis of local electronic properties in regions with internal stress near the dislocation lines. In our recent paper \cite{Putilov-JETPLett-19} we briefly reported on a \emph{single} observation of the line, outgoing from the center of the screw dislocation and therefore this line can be viewed as a hidden part of the dislocation loop. Here, we focus on the seeking of closely positioned screw dislocations in order (i) to confirm a possibility to visualize the hidden parts of the dislocation loops, and (ii) to realize whether two screw dislocations could be connected to each other by a hidden dislocation loop. The presented method of visualization of dislocation lines by STM/STS may be of interest for further investigations of dislocation-induced superconductivity in hybrid structures \cite{Fogel-PRL-01,Li-Nano-17}.

\section{Methods}

Experimental investigations of structural and electronic properties of quasi-two-dimensional Pb islands were carried out on an ultra-high vacuum (UHV) low-temperature scanning probe microscopy setup by Omicron Nanotechnology GmbH (base vacuum $2\cdot10^{-10}\,$mbar). Thermal deposition of Pb on the reconstructed surface Si(111)7$\times$7 was performed  at room temperature {\it in-situ} at pressure about $6\cdot10^{-10}\,$mbar. The orientation of the atomically flat terraces at the upper surface of the Pb islands corresponds to the (111) plane\cite{Altfeder-PRL-97,Altfeder-PRL-98,Su-PRL-01,Eom-PRL-06}. All STM/STS measurements were carried out at liquid nitrogen temperatures (from 78 to 81\,K) with electrochemically etched W tips additionally cleaned \emph{in-situ} by electron bombardment.

As it is generally accepted in scanning tunneling microscopy, the signal of the feedback loop acquired in the regime of the constant tunneling current is usually interpreted as a topography map $z=z(x,y)$. Since the presentation of the topography data requires additional procedures of image corrections, it is useful to plot unprocessed maps of the spatial variations of the tunneling current $I=I(x,y)$. Because of the unavoidable delay of the feedback loop, substantial deviations of the local current from the set-point value $\langle I\rangle$ occur near monatomic steps on the sample surface, point-like defects and screw dislocations, helping us to visualize surface defects directly during STM/STS measurements.

The local electronic properties of the Pb islands were studied by low-temperature STS in the regime with an active feedback, constant tunneling current $I$ and variable distance $z$ between the tip and the sample surface \cite{Schouteden-PRL-09,Aladyshkin-JPCM-20}. During the scanning process, the potential of the sample was harmonically modulated, containing dc-component $U^{\,}_0$ and ac-component $U^{\,}_1\cos(2\pi f^{\,}_0t)$, where $f^{\,}_0=7285$~Hz and $U^{\,}_1=40$\,mV. The amplitude of the first Fourier component of tunneling current $I^{\,}_1$ at the modulation frequency was routinely measured by the lock-in amplifier (Stanford Research, model SR\,830). The output voltage signal from the lock-in amplifier (proportional to $I^{\,}_1$) is then applied to the input of a built-in auxiliary analog-to-digital converter and recorded as $U^{\,}_{aux}$ in our measurement system. The measured $U^{\,}_{aux}$ value characterizes the tunneling conductance of the system 'sample--tip' at the energy $|e|U^{\,}_0$ above the Fermi energy $E^{\,}_F$. Hereafter we will use the term 'differential tunneling conductance' for the referring to $U^{\,}_{aux}$ for the sake of simplicity. Indeed, the positions of the local maxima on the dependence of $U^{\,}_{aux}$ on $U^{\,}_0$ should be close to that for the dependence of $dI/dU$ on $U^{\,}_0$, acquired in the regime of constant height (at least for low bias voltage). The absolute value of $U^{\,}_{aux}$ seems to be a non-informative parameter, since it depends on sensitivity and the bandwidth of the lock-in amplifier via a gain factor and time constant. However, spatial and bias-induced variations of $U^{\,}_{aux}$ measured at other fixed parameters are of particular interest because it makes possible to study spatial and energy dependence of the local differential tunneling conductance of the sample at a given energy.

\section{Results and Discussion}

We would like to describe local structural and electronic properties of three samples (S1, S2 and S3) --- the areas on the surface of three different Pb(111) islands with multiple screw dislocations of various types. Screw and edge dislocations typically appear in rather thick Pb islands and therefore their formation can be considered as a way of relaxing internal elastic deformation. All these samples were grown in similar conditions and have the same local thickness in the region of interest (about 9\,nm  with respect to the Si(111) surface). Data set for the sample S4 is presented in the Supporting Information.

\begin{figure*}[th!]
\centering{\includegraphics[width=10cm]{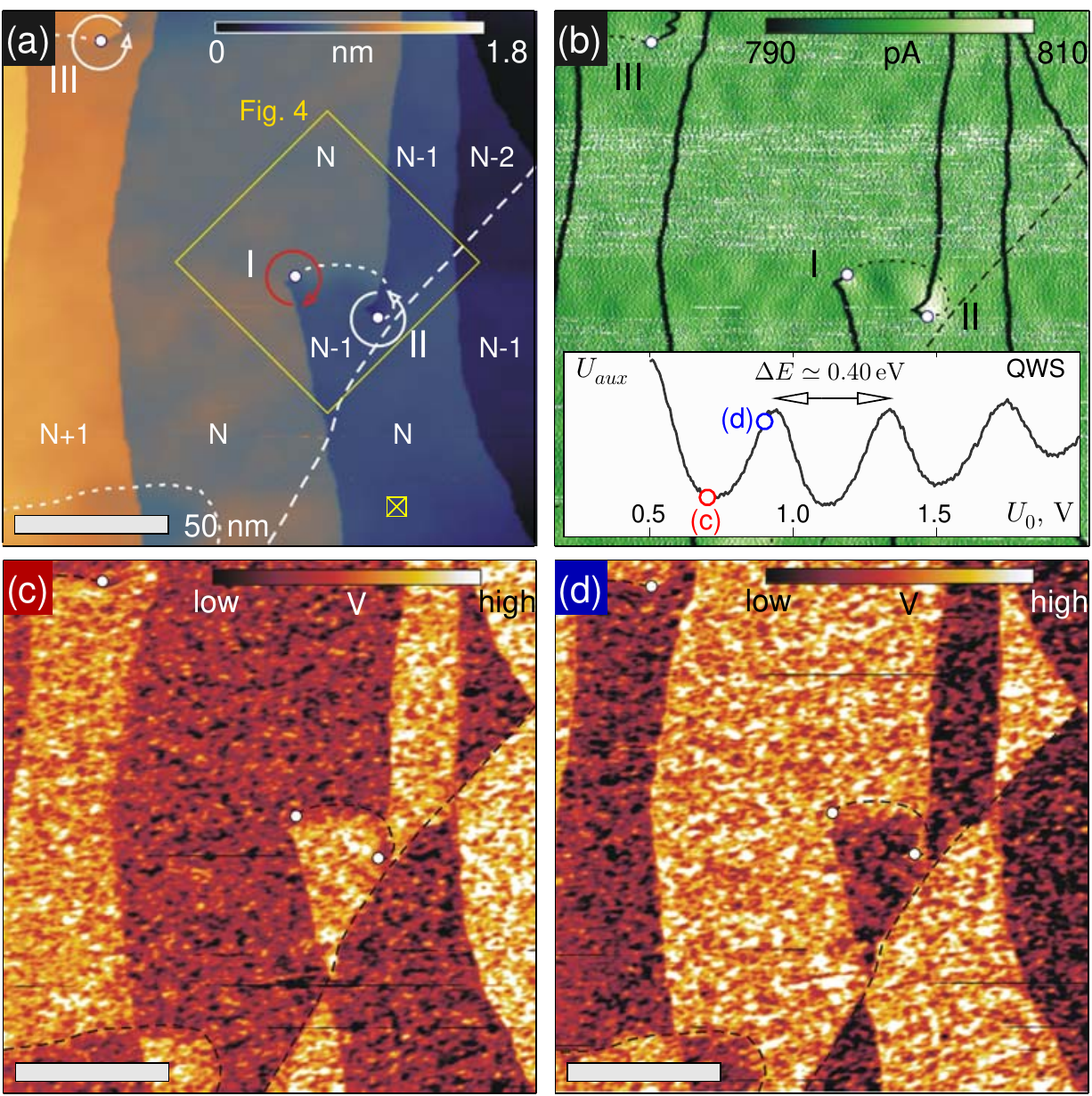}}
\caption{({\bf a}--{\bf d}) The aligned topographic image $z(x,y)$ (a),  the maps of the tunneling current $I(x,y)$ (b) and the differential tunneling conductance $U^{\,}_{aux}(x,y)$ (c, d), acquired simultaneously for sample S1 (image size is $174\times 174$\,nm$^2$, $U^{\,}_0=0.7$\,~V (a--c) and 0.9\,V (d), $\langle I\rangle=800\,$pA). All images (b)--(d) are raw data. The numbers in the panel (a) correspond to the thickness of the terraces in the $d^{\,}_{ML}$ units ($N\simeq 33$). White circles indicate the positions of the centers of the screw dislocations I--III. Dashed line shows the projection of the hidden monatomic step in the Si(111) substrate on the sample surface. Dotted lines show the projections of the hidden dislocation loops on the sample surface. The inset in panel (b) shows the dependence of $U^{\,}_{aux}$ on $U^{\,}_{0}$ at the point $\boxtimes$.
\label{Fig03}}
\end{figure*}

\vspace*{2mm}

\begin{figure*}[th!]
\centering{\includegraphics[width=10cm]{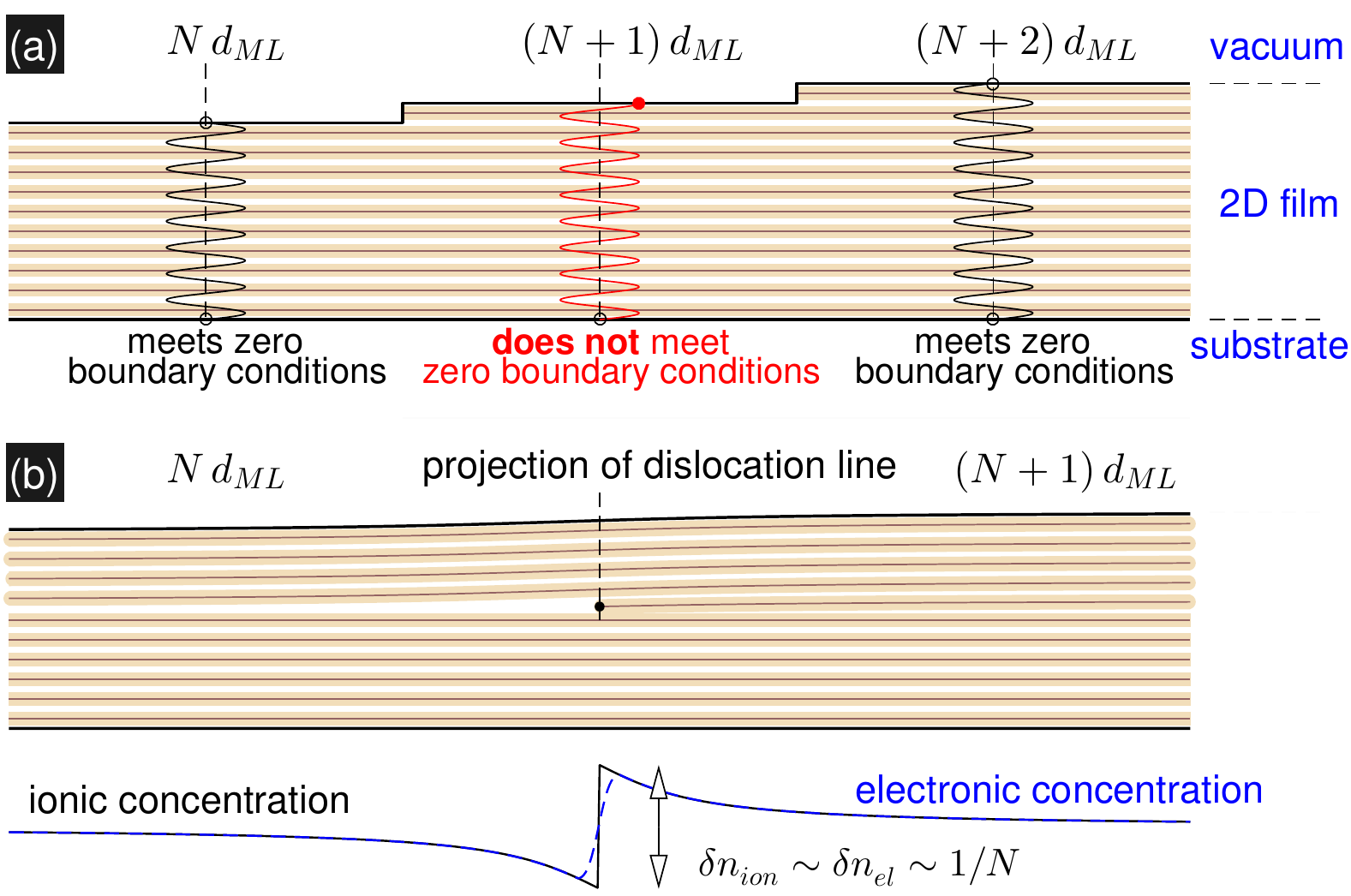}}
\caption{({\bf a}) Schematic presentation of the standing electronic waves with wavelength $\lambda=(4/3)\,d^{\,}_{ML}$, appearing in thin Pb film with few terraces of the quantized height. Assuming the infinite potential outside the film, one can conclude that the solution for the terrace with $N+1$ monolayers cannot be realized at the same energy $E$ as for the terraces with $N$ and $N+2$ monolayers. {\bf (b)}  Schematic presentation of the edge dislocation between the terraces with $N$ and $N+1$ monolayers as well as the expected profiles of ionic and electronic concentration near the dislocation line. }
\label{Fig02}
\end{figure*}

\noindent{\bf Sample S1.} The topographical image and the typical spectrum of the differential tunneling conductance are presented in Fig.~\ref{Fig03} (panels a and b). The pronounced oscillations of $U^{\,}_{aux}$ as a function of $U^{\,}_{0}$ point to the coherent resonant tunneling through the quantum-well states in a thin metallic film \cite{Altfeder-PRL-97,Altfeder-PRL-98,Su-PRL-01,Eom-PRL-06,Hong-PRB-09,Kim-SurfSci-15,Ustavshchikov-JETPLett-17,Putilov-JETPLett-19} (see schematics in Figs.~\ref{Fig01}a and \ref{Fig02}a). The quantum-size oscillations of the tunneling conductance can be easily detected in Pb(111) films up to 50 monolayers at liquid nitrogen temperatures \cite{Aladyshkin-JPCM-20} and up to 100 monolayers at sub-Kelvin temperatures \cite{Moore-2015}. It is obvious that the energy of electron localized in one-dimensional infinite potential well is equal to $E^{\,}_n \simeq E^{\,}_c + \hbar^2k^2_{\perp,n}/2m^*$, where $E^{\,}_c$ is the energy of the bottom of the conduction band in the free-electron-gas approximation; $m^*$ is the effective mass of electron in a crystalline sample; $n=1, 2\ldots$ is the integer index, $k^{\,}_{\perp,n}=\pi n/D$ is the allowed values of the transversed wave vector, and $D$ is the width of the quantum well ({\it i.e.} film thickness). If $E_n$ is close to $E^{\,}_F$, the above expression  can be expanded into a Taylor series with respect to $k^{\,}_n-k^{\,}_F$ as follows
	    \begin{equation}
   	    E^{\,}_n \simeq E^{\,}_F + \hbar v^{\,}_F \cdot \left(\frac{\pi n}{D} - k^{\,}_F\right),
        \label{Eq1}
    	\end{equation}
where $E^{\,}_c = E^{\,}_F - \hbar^2k^2_{F}/2m^*$;  $k^{\,}_F$ and $v^{\,}_F=\hbar k^{\,}_F/m^*$ are the Fermi momentum and the Fermi velocity. Taking the expected period of the quantum-size oscillations $\Delta E\equiv E^{\,}_{n+1}-E^{\,}_n=\pi\hbar v^{\,}_F/D$ from Eq.~(\ref{Eq1}) and the observed period $\Delta E\simeq 0.40\,$eV from the inset in Fig.~\ref{Fig03}b, one can estimate the local thickness $D\simeq 9.3\,$nm with respect to the Si(111) surface using the relationship \cite{Altfeder-PRL-97} $D\simeq \pi\hbar v^{\,}_F/\Delta E$, where $v^{\,}_F\simeq 1.8\cdot10^{8}$\,cm/s is the Fermi velocity for the Pb(111) films \cite{Altfeder-PRL-97,Ustavshchikov-JETPLett-17}.

The color-coded maps of the differential tunneling conductance at two different energies ($|e|U^{\,}_0=0.7$ and 0.9\,eV), acquired simultaneously with the topographic image at forward and backward scanning directions are presented in Fig.~\ref{Fig03} (panels c and d). The interpretation of these maps is based on the concept of commensurability between the Fermi wave length $\lambda^{\,}_F\simeq 0.394\,$nm (Refs. \cite{Su-PRL-01,Eom-PRL-06,Hong-PRB-09}) and the thickness of the Pb(111) monolayer $d^{\,}_{ML}=0.286\,$nm. By substituting $D=N\,d^{\,}_{ML}$ and $k^{\,}_F=2\pi/\lambda^{\,}_F$, we rewrite Eq.~(\ref{Eq1}) in the following form
	\begin{eqnarray}
 	\nonumber
	E^{(N)}_{\,\,n} \simeq E^{\,}_F + \frac{\pi\hbar v^{\,}_F}{N\,d^{\,}_{ML}}\cdot\left(n - 2N\,\frac{d^{\,}_{ML}}{\lambda^{\,}_F}\right),
	\end{eqnarray}
where $n$ is the number of the half-waves and $N$ is the number of monolayers.  The energy of the standing wave with $n+3$ half-waves in the film with $N+2$ monolayers is equal to
	\begin{eqnarray}
	\nonumber
	E^{(N+2)}_{\,\,n+3} \simeq E^{\,}_F + \frac{\pi\hbar v^{\,}_F}{(N+2)\,d^{\,}_{ML}}\cdot \left((n+3) -
    2(N+2)\,\frac{d^{\,}_{ML}}{\lambda^{\,}_F}\right).
	\end{eqnarray}
Because $d^{\,}_{ML}/\lambda^{\,}_F\simeq 3/4$ for the Pb films, it is easy to see that $(n+3) - 2(N+2)\,d^{\,}_{ML}/\lambda^{\,}_F=n - 2N\,d^{\,}_{ML}/\lambda^{\,}_F$ and $E^{(N)}_{\,\,n}\simeq E^{(N+2)}_{\,\,n+3}$ provided that  $E^{\,}_n\simeq E^{\,}_F$ and $N\gg 1$. As a consequence, the Pb(111) terraces of the same parity of $N$ should have almost equal tunneling conductance with the magnitude depending on the bias voltage. Since the differential conductance at the location $\boxtimes$ has local maximum at 0.9\,V and $N=D/d^{\,}_{ML}\simeq 33$, all bright areas in Fig.~\ref{Fig01}d should correspond to the terraces with odd numbers of the Pb monolayers and vice versa \cite{Ustavshchikov-JETPLett-17,Putilov-JETPLett-19}. At the energy $|e|U^{\,}_0=0.7$\,eV the spatial distribution of bright and dark areas become inverted with respect to that for $|e|U^{\,}_0=0.9$\,eV (compare panels c and d in Fig.~\ref{Fig03}). Small-scale modulation of the differential conductance at the atomically flat terraces apparently results from the periodic modulation of the buried Si(111)7$\times$7 surface \cite{Altfeder-PRL-98}.

\begin{figure*}[th!]
\centering{\includegraphics[width=16cm]{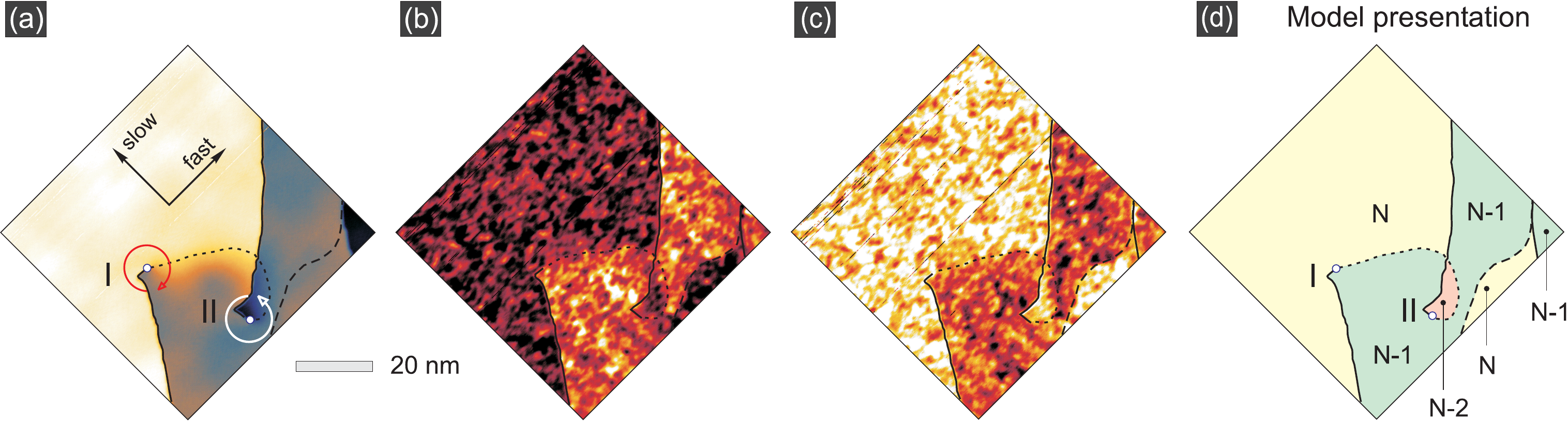}}
\caption{(color online) ({\bf a}--{\bf c}) Aligned topography map $z(x,y)$ and two maps of the differential tunneling conductance $U^{\,}_{aux}(x,y)$ (b, c) for the area on the surface of sample S1 near screw dislocations I and II (image size is $69.6\times 69.6$\,nm$^2$, $U^{\,}_0=0.7$\,V (a, b) and $0.9$\,V (c), $\langle I\rangle=800~$pA). The arrows in the panel (a) show the directions of fast and slow scanning.  ({\bf d}) Schematic presentation of the projection of the hidden part of the dislocation loop between screw dislocations I and II (dotted line), the invisible monatomic step in the substrate (dashed line) as well as visible monatomic steps at the upper surface (solid lines) for the square-shaped area in Fig.~\ref{Fig03}a.}
\label{Fig04}
\end{figure*}

The presence of the screw dislocations at the upper surface expectedly leads to a gradual variation in the visible height of the Pb terraces. In particular, the local thickness decreases at one Pb monolayer along the paths traversed clockwise for the screw dislocations I and counter-clockwise for the screw dislocations II and III (see red and white round arrows in Fig.~\ref{Fig03}a). However, the variation in the film thickness around the centers of the screw dislocation does not result in a gradual change of the differential tunneling conductance: the conductance changes drastically upon crossing a line invisible at the topographic image. The line connecting the centers of dislocations I and II can be considered as a projection of the hidden part of the dislocation loop inside the film between two screw dislocations with the opposite Burgers vectors. The length of this projection is about 30 nm, what is approximately three times larger than the local film thickness.

Because the projection of this dislocation line is almost parallel to the direction of fast scanning; for the detailed analysis we repeat the STS measurements for the square-shaped area rotated at 45$^{\circ}$ (Fig.~\ref{Fig04}) in order to demonstrate substantially different widths of the transient areas in the topography near monatomic steps at the upper surface and the hidden dislocation loop. A model presentation of the mutual arrangement of the monatomic steps, the screw dislocations and the hidden dislocation loop is shown in Fig.~\ref{Fig04}d.

We would like to present the qualitative interpretation of a possibility to detect the hidden dislocations by means of STS \cite{Putilov-JETPLett-19}. In the vicinity of the edge dislocation, the numbers of Pb monolayers $N(x)$ should change at one while the local film thickness $D(x)$ continuously varies at $d^{\,}_{ML} $ at length scales of the order of 10--20 nm (see schematics in Fig.\,\ref{Fig02}b). Assuming uniform distribution of the material parameters in the lateral direction, one can conclude that the ionic concentration $n^{\,}_{ion}$ being proportional to $N(x)/D(x)$ should have a jump in the concentration  of the order of $1/(N\,d^{\,}_{ML})$. The electronic concentration $n^{\,}_{el}$ should be also nonuniform and thus differ from the equilibrium value $n^{(0)}_{el}$ for atomically flat terraces. This problem of the spatial distribution of electron density deserves detailed treatment and it is beyond this paper. The jump in the $n^{\,}_{el}$ value for $N\gg 1$ can be estimated as $\delta n^{\,}_{el} \simeq n^{(0)}_{el}/N$. Since the Fermi energy in the free-electron-gas approximation depends on the electronic concentration as $E^{,}_F =(\hbar^2/2m^*)\,(3\pi^2\,n_{el})^{2/3}$  \cite{Ashcroft-Mermin}, one can expect that the jump in $\delta n^{\,}_{el}$ should produce the increase in $E^{\,}_F$ at the value of the order of
    \begin{eqnarray}
    \nonumber
    \delta E^{\,}_F \simeq \left(\frac{\hbar^2}{2m^*}\right)\,\left(3\pi^2\right)^{2/3}\,\left(n^{(0)}_{el}\right)^{2/3}\times \left\{\left(1+\frac{1}{N}\right)^{2/3} - 1\right\} \simeq \frac{2}{3}\,\frac{E^{(0)}_F}{N}
    \label{Eq3}
    \end{eqnarray}
for $N\gg 1$. Thus the bottom of the conduction band in the area with larger $N$ should be shifted down at $\delta E^{\,}_c\sim \delta E^{\,}_F$ to equalize the level of the electro-chemical potential in thin metallic film of variable thickness. Taking $E^{\,}_F=9.47$\,eV for bulk Pb  \cite{Ashcroft-Mermin} and $N\sim 33$, one can get the estimate $\delta E^{\,}_c \sim 190\,$meV. Since $\delta E^{\,}_c$ is close to the half of the period $\Delta E$ of the quantum-size oscillations, the relocation of the tip across the invisible part of the dislocation loop should be accompanied by a sharp change in brightness for the maps of the tunneling conductance.

\begin{figure*}[th!]
\centering{\includegraphics[width=15.5cm]{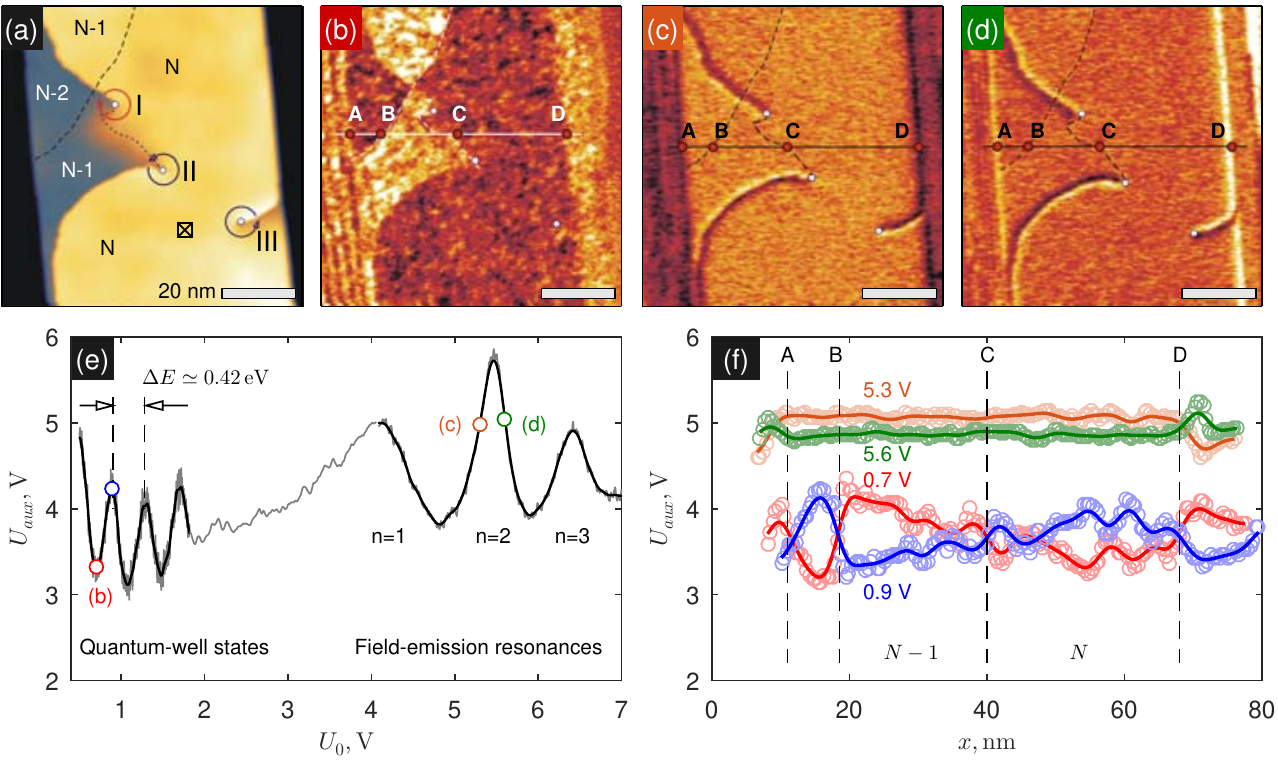}}
\caption{({\bf a}--{\bf d})  Aligned topographic image $z(x,y)$ (a) and the $U^{\,}_{aux}(x,y)$ maps in the tunneling (b) and the field-emission (c,d) regimes for sample S2 (image size is $81.2\times 81.2$~nm$^2$, $U^{\,}_0=0.7$\,V (a,b), 5.3\,V (c) and 5.6\,V (d), $\langle I\rangle=200~$pA). The numbers in panel (a) correspond to the local thickness of each terrace in the $d^{,}_{ML}$ units ($N\simeq 31$). White circles indicate the positions of the centers of screw dislocations I--III. A dashed line shows the projection of the hidden monatomic step in the Si(111) substrate on the upper terrace, the part of this line below the bunch of the monatomic steps at the upper surface is not shown because of uncertainty of its position. A dotted line shows the projection of the hidden dislocation loop between the dislocations I--II on the upper terrace.  ({\bf e}) Dependence of $U^{\,}_{aux}$ on $U^{\,}_{0}$ recorded at the point $\boxtimes$.  ({\bf f}) Spatial variations of $U^{\,}_{aux}$ along the lines $A-D$. All solid lines are guides for the eyes obtained by a running Gaussian averaging of the measured values within a window of 1\,nm.
\label{Fig05}}
\end{figure*}



\vspace*{2mm}

\noindent{\bf Sample S2.} Fig.~\ref{Fig05} shows topographical image (a) and three maps of the differential tunneling conductance at $|e|U^{\,}_0=0.7\,$eV (b), 5.3\,eV (c) and 5.6\,eV (d). The local spectrum of the differential tunneling conductance (panel e) makes it possible to estimate the period of the quantum-size oscillations ($\Delta E\simeq 0.42\,$eV) and then the local thickness: $D\simeq 8.9\,$nm or $N=D/d^{\,}_{ML}\simeq 31$. Similar to the data shown in Fig.~\ref{Fig03}c, all bright and dark areas in Fig.~\ref{Fig05}b corresponds to the terraces with even and odd numbers of the Pb monolayers, respectively.

Considering the distribution of intensities on the $U^{\,}_{aux}(x,y)$ map at a low bias voltage, we can easily recognize the hidden monatomic step in the Si substrate and the dislocation loop, connecting the screw dislocations I and II with the opposite Burgers vectors. It is interesting that there is no line outcoming from the center of the screw dislocation III. A probable interpretation is that this dislocation line is oriented perpendicularly to the film surface without noticeable projection to the sample surface. The most plausible explanation is that the cluster of closely positioned point defects like vacancies on the surface of the substrate can produce vertically oriented screw dislocation running across the film. The screw dislocations in thick Pb films help to decrease the energy of elastic deformations both in the bulk of the film and at the film-substrate interface. The vertical orientation of such the dislocation apparently correspond to its minimal length. In any case, our method makes it possible to detect the dislocations with different internal structure.

The spatial dependences of $U^{\,}_{aux}$ on the $x-$coordinate along the line $A-D$ running exactly between the centers of dislocations I and II at $|e|U^{\,}_0=0.7$ and 0.9\,eV are displayed in Fig.~\ref{Fig05}f. We introduce points $A$ and $D$ as the intersection points of this median line and the edges of the upper terrace. These points can be unequivocally determined by the abrupt changes in the lock-in signal at all voltages (see panels b--d). We use this definition in order to remove the creep of a piezo-scanner and thus to combine the different cross-sectional views from the two-dimensional maps $U^{\,}_{aux}(x,y)$ acquired serially at different energies. The points $B$ and $C$ are the positions of the monatomic step in the substrate and the hidden part of the dislocation loop, correspondingly, and therefore visible only at low bias values. The pronounced anti-correlated variations of $U^{\,}_{aux}(x)$ measured at $|e|U^{\,}_0=0.7$ and 0.9\,eV evidence for the changes in the local height of the film near these points $A$, $B$, $C$ and $D$.

In addition to the STS measurements at a low bias voltage we study the dependence of $U^{\,}_{aux}$ on $U^{\,}_0$ in the field-emission regime at a high bias voltage (Fig.~\ref{Fig05}e). Aperiodic large-scale oscillations of the differential conductance, also called as the Gundlach oscillations \cite{Gundlach-SolStateElectr-66} or field-emission resonances (FER), correspond to the resonant tunneling through quasi-stationary states localized above the surface of the sample (see schematics in Fig.~\ref{Fig01}b). The elementary theory based on the quasi-classical approximation for the electron trapped in the triangular potential well returns the following expression\cite{Aladyshkin-JPCM-20} for the bias voltage $U^{(n)}$, corresponding to the $n-$th FER peak
    \begin{eqnarray}
    \label{Eq:IPS-triangular-well-3}
    |e|U^{(n)} \simeq W^{\,}_s + \left(\frac{3}{2}\,\frac{\pi\hbar}{\sqrt{2m^{\,}_{0}}}\right)^{2/3}\cdot F^{*\,2/3}_n\cdot\left(n-\frac{1}{4}\right)^{2/3}.
    \end{eqnarray}
Here,  $n = 1, 2 \ldots$; $F^*=(|e|U^{(n)} + W^{\,}_t-W^{\,}_s)/h$ is the gradient of the potential energy, proportional to the local electric field;  $W^{\,}_s$ and $W^{\,}_t$ are the work functions for the sample and the STM tip; and $h$ is the distance between the tip and the surface. The expression (\ref{Eq:IPS-triangular-well-3}) is identical to the relationship derived for the positions of the $dI/dU$ maxima for the tunnelling junction with a trapezoidal potential barrier between two metals in the free-electron-gas approximation \cite{Gundlach-SolStateElectr-66,Kolesnychenko-PhysB-00}.  It is important to note that the analysis of the field emission resonances makes it possible to detect the spatial variations of the local work function and surface potential \cite{Rienks-PRB-05,Pivetta-PRB-05,Ploigt-PRB-07,Lin-PRL-07,Ruffieux-PRL-09}.

The single-point spectroscopic measurement acquired at sweeping mean bias potential potentially provides more information related to the local electronic properties of the studied system \cite{Rienks-PRB-05,Pivetta-PRB-05,Ploigt-PRB-07,Lin-PRL-07,Ruffieux-PRL-09,Borca-2D-20}. At liquid nitrogen temperatures the creep of a piezo-scanner is noticeable and the STM tip uncontrollably shifts from the measurement position during the acquiring the single-point spectroscopic dependence. In order to resolve a conflict between the energetic and spatial resolutions, we perform the STS measurements at  fixed energy with a high spatial resolution. One can expect that the variations of the surface potential and/or local work function will induce the coherent shift of all field-emission resonances. The second FER peaks positioned at 5.45\,eV has the maximum amplitude and therefore it is characterized by the maximal slope $|dU^{\,}_{aux}/dU^{\,}_0|$ at left and right sides of this peak (Fig.~\ref{Fig05}e), therefore a tiny displacement of the second field-emission peak should cause the detectable variations of $U^{\,}_{aux}$. Thus, it is interesting to test the local variations of the work function and electrical potential near the screw dislocations by measuring the spatial dependence of $U^{\,}_{aux}$ on the lateral coordinates $x$ and $y$ provided that the mean potential $U^{\,}_0$ is close to the $U^{(2)}$ value.

The $U^{\,}_{aux}(x,y)$ maps acquired at $U^{\,}_0=5.3$ and 5.6\,V are shown in Fig.~\ref{Fig05}c,d. We emphasize that there are no systematic variations of $U^{\,}_{aux}$ on the $x$ coordinate along the $A-D$ line running between two screw dislocations I and II  (Fig.~\ref{Fig05}f). We consider the invariability of the second FER as an {\it indication} of constancy of the electro-chemical potential near the $B$ and $C$ points. It is worth noting that the $U^{\,}_{aux}$ signal, measured in the field-emission regime ($U^{\,}_0\gtrsim 4.5\,$V), is strongly enhanced/suppressed near the monatomic step on the upper surface of the Pb film in contract to that for the tunneling regime ($U^{\,}_0\lesssim 2\,$V). This can be considered as a strong argument in favor of applicability and sensitivity of this method. The bias-dependent evolution of $U^{\,}_{aux}$ can be described if one assumes that the FER peaks shifts to the higher energies for a thicker part of the monatomic step in the Pb film and vice versa. The detailed experimental investigations of the effects of the local electric potential and local work function near monatomic steps and hidden dislocation lines are beyond the aim of this paper.

\begin{figure*}[t!]
\centering{\includegraphics[width=16cm]{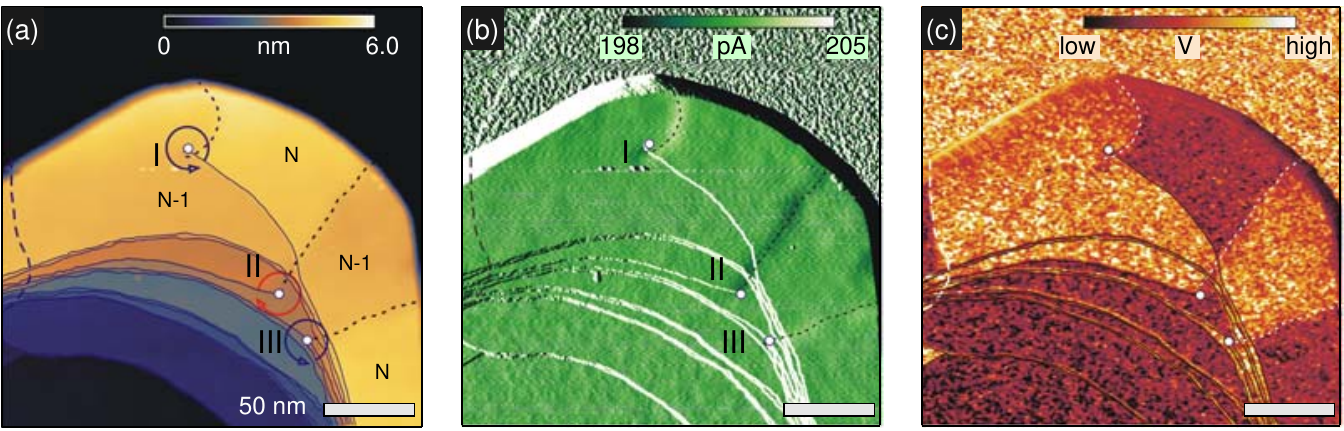}}
\caption{({\bf a}--{\bf c}) Aligned topographic image $z(x,y)$ (a),  the maps of the tunneling current  $I(x,y)$ (b), and differential tunneling conductance $U^{\,}_{aux}(x,y)$ (c) acquired simultaneously for sample S3 (image size is $232\times 232$~nm$^2$, $U^{\,}_0=0.7$~V, $\langle I\rangle=400~$pA). The numbers in panel (a) correspond to the local thickness of the terrace in the $d^{\,}_{ML}$ units ($N\simeq 31$). White circles mark the position of the screw dislocations, solid blue lines depict the position of the monatomic steps at the upper surface. A dashed line shows the projection of the hidden monatomic step in the Si(111) substrate on the sample surface. Dotted lines show the projections of the hidden dislocation loops on the sample surface.
\label{Fig06}}
\end{figure*}

\vspace*{2mm}

\noindent{\bf Sample S3.} Fig.~\ref{Fig06} shows the topographical image, the map of the tunneling current and the map of the differential tunneling conductance at $|e|U^{\,}_0=0.7\,$eV. It is easy to see that the edge dislocations generated by the screw dislocations with the opposite Burgers vectors can leave the Pb island instead of being connected to each other. Interestingly, the projection of the hidden part of the dislocation line and the monatomic steps at the upper surface can intersect thus indicating that the edge dislocation propagates into the bulk below one and few Pb monolayers like a cable running under a carpet.

We would like to discuss the effect of the scan speed on the visibility of the dislocation lines for the given loop gain (typically 1.5--2\%). The scan speed varies from 15\,nm/s (Figs.~\ref{Fig03}--\ref{Fig05}) to 38.5\,nm/s for the data in Fig.~\ref{Fig06}. Such a two-fold increase in the scan speed does not substantially worsen the spatial resolution for the maps of the differential tunneling conductance provided that the time constant of the lock-in amplifier is properly adjusted. However this enhances the standard deviation of the instant current from the set-point value (compare limiting values in Figs.~\ref{Fig03}b and \ref{Fig06}b). Indeed, the higher the speed of scanning, the less the accuracy of the current stabilization is. It eventually leads to the fact that the hidden parts of the dislocation lines become visible in the map of the current variations (Fig.~\ref{Fig06}b). Such controlled imperfection of the feedback-loop performance gives us an independent channel for the visualization of the areas with non-quantized height variations directly during the measurement process even without a lock-in amplifier.

\section{Conclusion}

We experimentally investigated the local electronic properties of quasi-two-dimensional Pb islands grown on the reconstructed Si(111)$7\times7$ surface by means of low-temperature scanning tunneling microscopy and spectroscopy. An important feature of the tested Pb islands was the presence of the screw dislocations of different types on the surface. By comparing the topography map and the maps of the differential tunneling conductance on different energies we clearly determined the positions of the projection of the hidden part of the dislocation line to the sample surface. We demonstrated that two closely-positioned screw dislocations with the opposite Burgers vectors can connect to each other by the dislocation loop below the surface. Alternatively, screw dislocations can produce independent edge dislocations running below the sample surface towards the edges of the Pb island. In both cases the hidden dislocation lines cause the non-quantized variation of the local thickness of the Pb terraces. We argued that the hidden dislocations lines can also be visible on the maps of the tunneling current what makes the visualization of such lines possible even without the technique of synchronous detection.

\section{Acknowlednements}

The authors thank V. S. Stolyarov and A. S. Mel'nikov for valuable comments. The work was performed with the use of the facilities at the Common Research Center 'Physics and Technology of Micro- and Nanostructures' at Institute for Physics of Microstructures RAS. The reported study was partly funded by the Russian Fund for Basic Research (No. 19-02-00528, STM-STS measurements), partly funded by the Russian State Contracts for Institute for Physics of Microstructures RAS (No. 0030-2021-0020, sample preparation) and for Osipyan Institute of Solid State Physics RAS (interpretation of results).

\newpage

\section*{\textcolor[rgb]{0.00,0.07,1.00}{SUPPORTING INFORMATION}}

\section*{Sample S2}

\begin{figure*}[ht!]
\centering{\includegraphics[width=14.5cm]{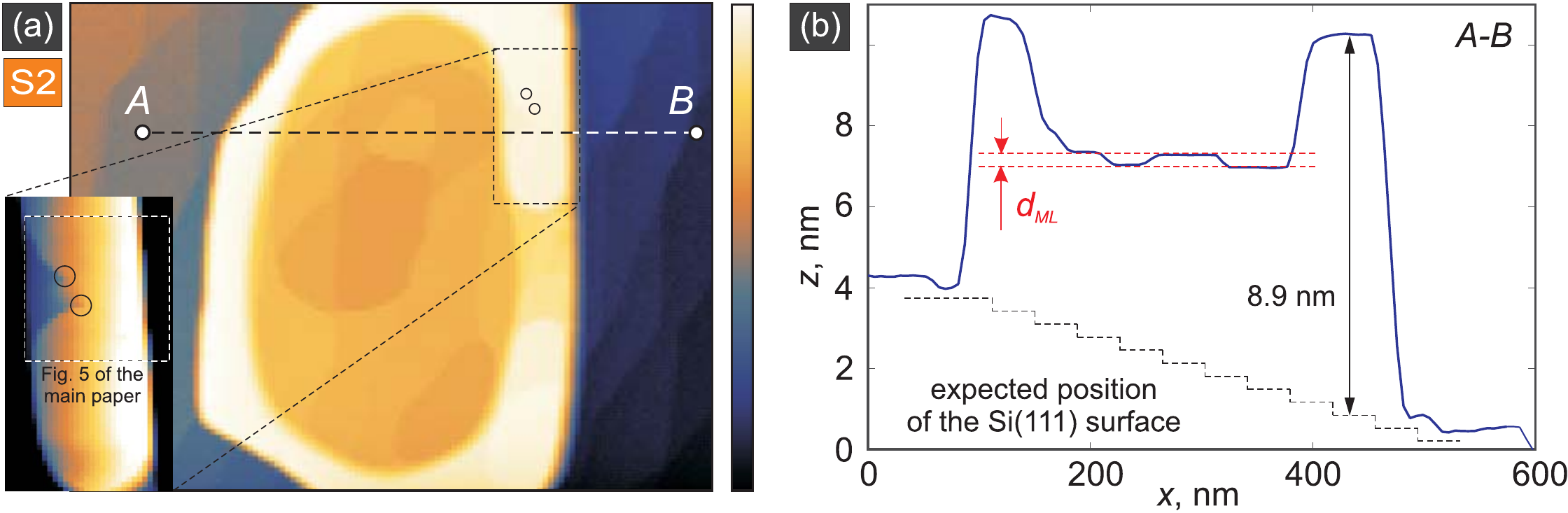}}
\caption{({\bf a}) Overview topography map for the sample S2 (image size is $696\times 522$~nm$^2$, $U^{\,}_0=0.5$~V, $\langle I\rangle=200~$pA); the inset shows the zoomed part of the Pb island with two screw dislocations marked by circles (image size $93\times162$~nm$^2$). ({\bf b}) The profile $z(x)$ along the $A-B$ line, depicted by the dashed line in the panel (a); the value $d^{\,}_{ML}=0.286\,$nm corresponds to the height of the monatomic step at the Pb(111) surface.}
\label{Fig-SM-02}
\end{figure*}

\begin{figure*}[ht!]
\centering{\includegraphics[width=13.5cm]{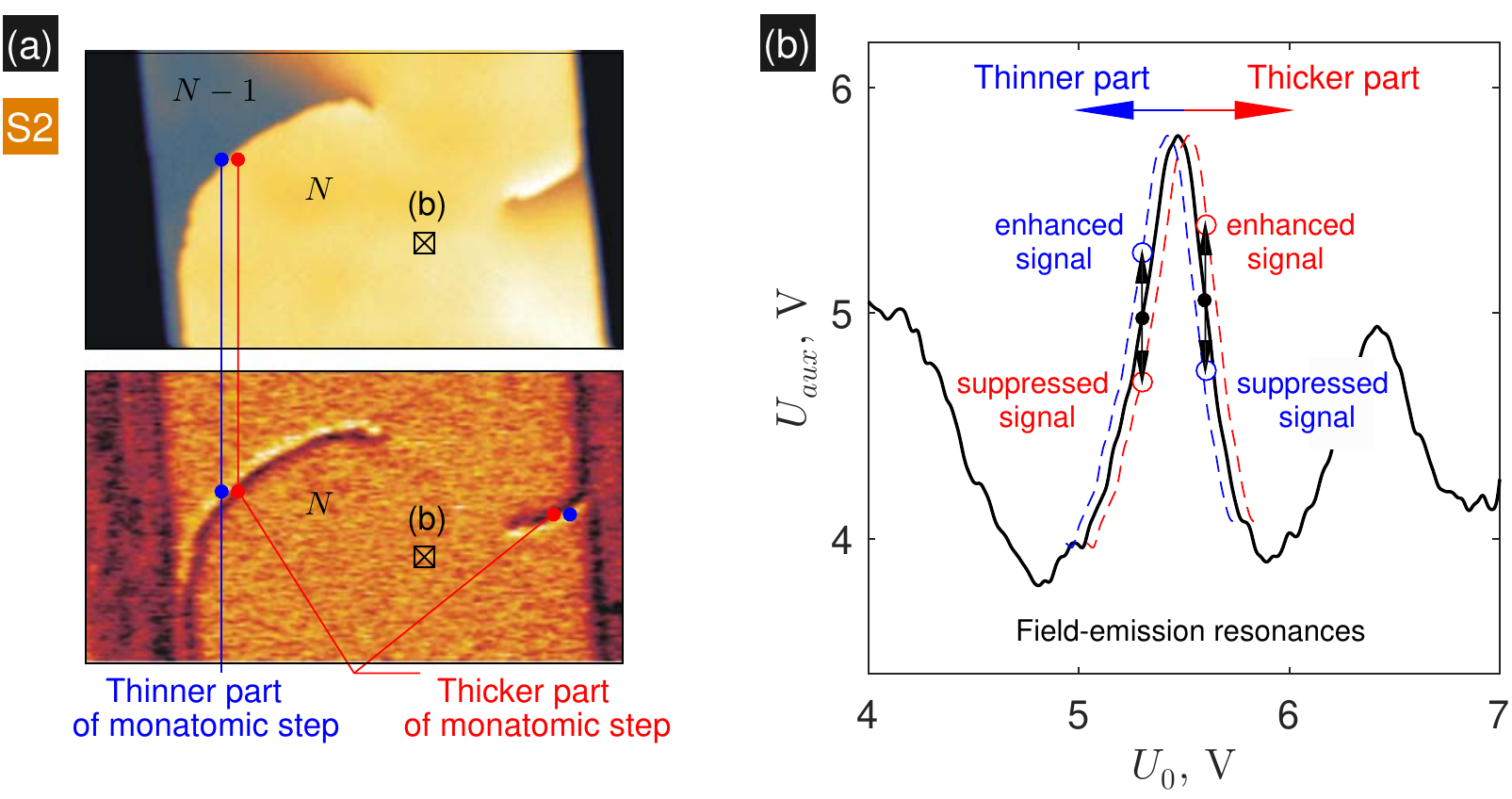}}
\caption{({\bf a}) Aligned topography map and the map of the differential tunneling conductance $U^{\,}_{aux}(x,y)$ for the sample S2 in the field-emission regime (image size is $81.2\times 47$~nm$^2$, $U^{\,}_0=5.3$~V, $\langle I\rangle=200~$pA, $N\simeq 31$). ({\bf b}) The dependence of $U^{\,}_{aux}$ on $U^{\,}_{0}$ recorded at the point $\boxtimes$ in the field-emission regime (black solid line). Dashed red and blue lines represent the expected transformation of the second field-emission resonance for thicker and thinner parts for the Pb film near the monatomic step at the upper surface, respectively. Vertical arrows describe the expected variations in the magnitude of the $U^{\,}_{aux}$ signal at the left and right sides of the resonance at 5.3 and 5.6\,V.}
\label{Fig-SM-02b}
\end{figure*}

\newpage
\section*{Sample S3}

\begin{figure*}[ht!]
\centering{\includegraphics[width=18cm]{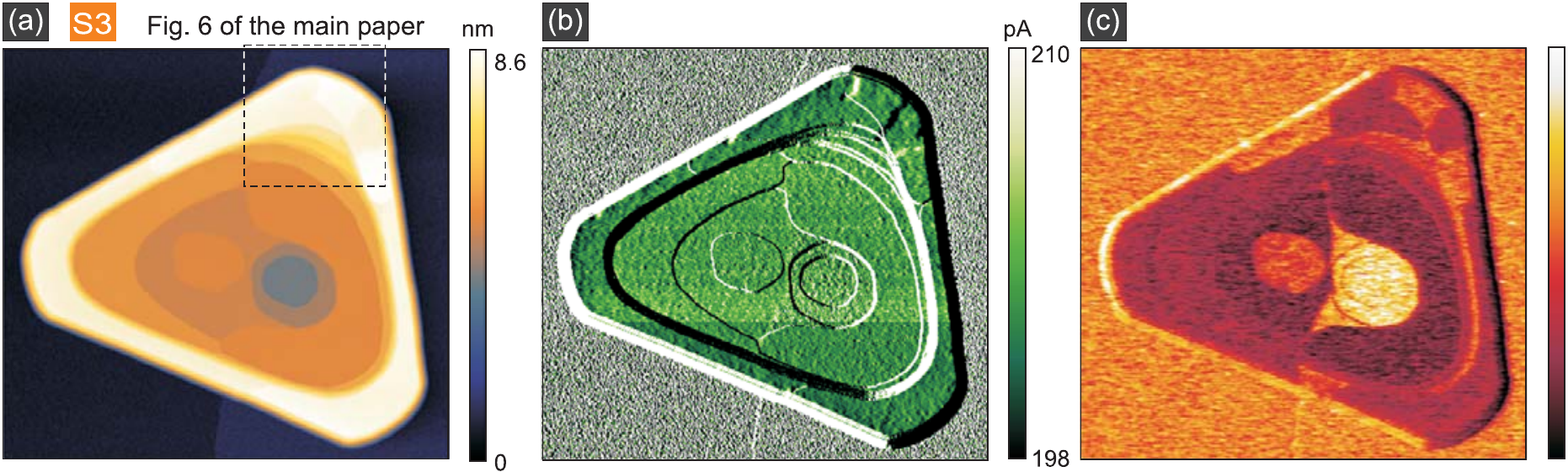}}
\caption{(color online) ({\bf a-c}) Aligned overview topography map (a), the map of the tunneling current variations $I(x,y)$ (b), and map of the differential tunneling conductance $U^{\,}_{aux}(x,y)$ (c) acquired simultaneously for the sample S3 (image size is $696\times 638$\,nm$^2$, $U^{\,}_0=0.5$\,V, $\langle I\rangle=200\,$pA). All images (b)--(c) are raw data.}
\label{Fig-SM-03}
\end{figure*}

\begin{figure*}[ht!]
\centering{\includegraphics[width=16cm]{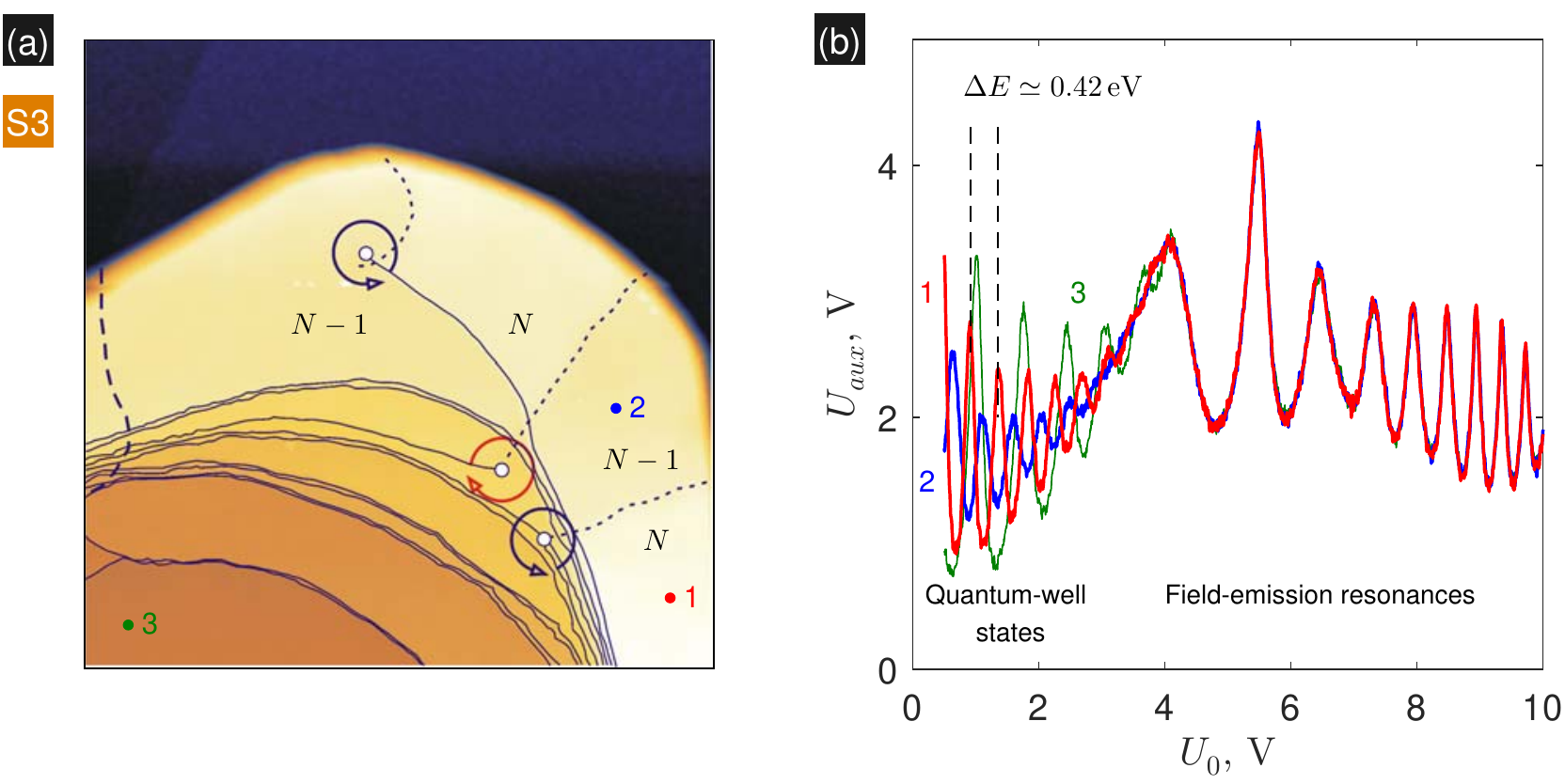}}
\caption{(color online)  ({\bf a}) Aligned topography map for the sample S3 near the edge of the island (image size is $232\times 232$\,nm$^2$, $U^{\,}_0=0.5$\,V, $\langle I\rangle=200\,$pA). Symbols $\circ$ mark the positions of the centers of the screw dislocations. Dashed line shows the projection of the hidden monatomic step in the Si(111) substrate on the sample surface. Dotted lines show the projection of the hidden dislocation loops on the sample surface. {\bf (b)} The local tunneling spectra -- the dependences of $U^{\,}_{aux}$ on $U^{\,}_{0}$ -- recorded at the points 1, 2 and 3 at $\langle I\rangle=400\,$pA. All spectral lines are raw data.}
\label{Fig-SM-04}
\end{figure*}


\newpage

\section*{Sample S4}

\begin{figure*}[ht!]
\centering{\includegraphics[width=12cm]{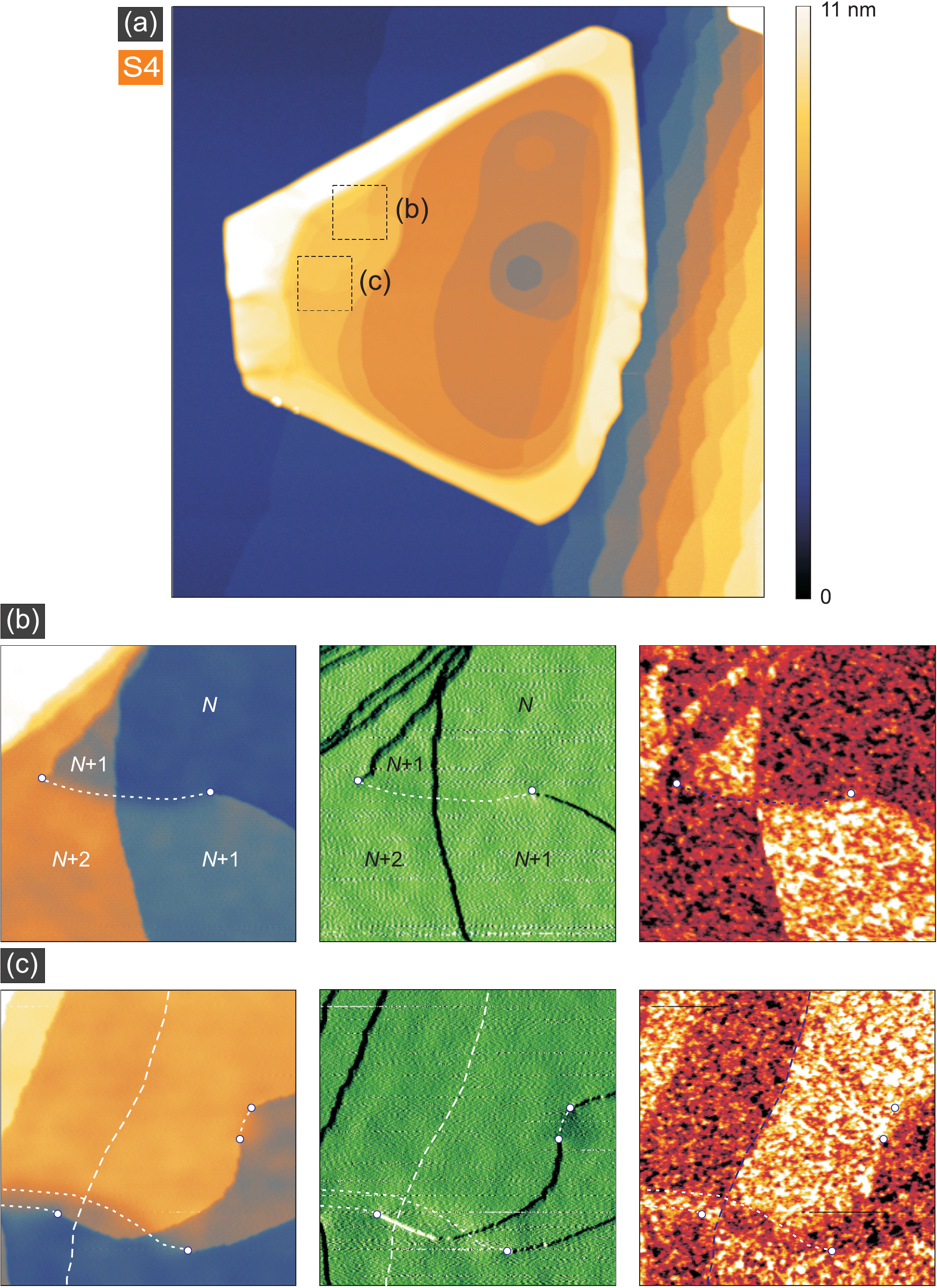}}
\caption{(color online) ({\bf a}) Aligned overview topography map for the sample S4 (image size is $1044\times 1044$\,nm$^2$, $U^{\,}_0=0.75$\,V, $\langle I\rangle=300\,$pA). {\bf (b, c)} Topography map as well as maps of the current variations and the differential tunneling conductance for the square-shaped areas $93\times 93$\,nm$^2$, depicted in the panel (a). Symbols $\circ$ mark the positions of the centers of the screw dislocations. Dashed line shows the projection of the hidden monatomic step in the Si(111) substrate on the sample surface. Dotted lines show the projection of the hidden dislocation loops on the sample surface.}
\label{Fig-SM-05}
\end{figure*}

\begin{figure*}[h!]
\centering{\includegraphics[width=17cm]{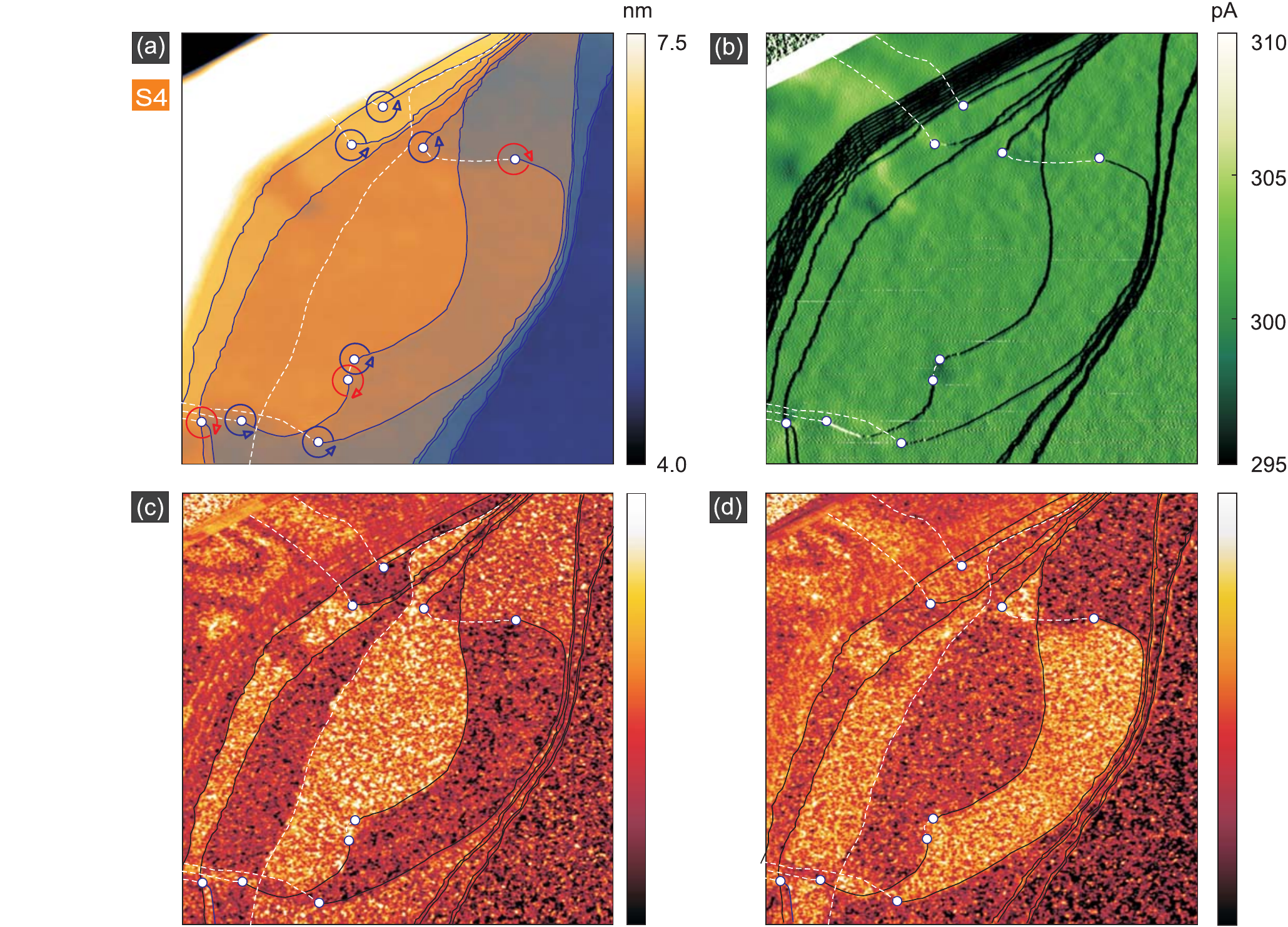}}
\caption{(color online) ({\bf a}--{\bf d}) Aligned topographic image $z(x,y)$ (a), the map of the tunneling current variations $I(x,y)$ (b), and two maps of the differential tunneling conductance $U^{\,}_{aux}(x,y)$ (c, d) acquired simultaneously for the inner part of the sample S4 (image size is $232\times 232$~nm$^2$, $U^{\,}_0=0.5$~V (a--c) and 0.6\,V (d), $\langle I\rangle=300~$pA). All images (b)--(d) are raw data. Symbols $\circ$ mark the positions of the centers of the screw dislocations. Dashed line shows the projection of the hidden monatomic step in the Si(111) substrate on the sample surface. Dotted lines show the projection of the hidden dislocation loops on the sample surface.}
\label{Fig07}
\end{figure*}

\end{document}